\documentclass[twocolumn,showpacs,preprintnumbers,amsmath,amssymb]{revtex4}

\usepackage{graphicx}
\usepackage{dcolumn}
\usepackage{bm}

\begin{document}

\title{Engineering local optimality in Quantum Monte Carlo algorithms}

\author{Lode Pollet}
\email{pollet@itp.phys.ethz.ch}
\affiliation{Theoretische Physik, \\ ETH Z{\"u}rich, CH - 8093 Z{\"u}rich, Switzerland}
\author{Kris Van Houcke}
\author{Stefan M. A. Rombouts}
\affiliation{Vakgroep subatomaire en stralingsfysica, \\
  Proeftuinstraat 86, 
   Universiteit Gent, Belgium}

\date{\today}

\begin{abstract}
Quantum Monte Carlo algorithms based on a world-line representation such as
the worm algorithm and the directed loop algorithm are among the most powerful
numerical techniques for the simulation of non-frustrated spin models and
of bosonic models. Both algorithms work in the grand-canonical ensemble and
can have a winding number larger than zero. However, they retain a lot of intrinsic
degrees of freedom which can be used to optimize the algorithm. We let us
guide by the rigorous statements on the globally optimal form of Markov chain
Monte Carlo simulations in order to devise a locally optimal formulation of
the worm algorithm while incorporating ideas from the directed loop
algorithm. We provide numerical examples for the soft-core Bose-Hubbard model
and various spin-$S$ models. 
\end{abstract}

\maketitle

\section{Introduction}
Monte Carlo methods have become a standard numerical tool in many branches of
science, offering exact results in a statistical sense~\cite{Liu01}. In
physics, some Monte Carlo
algorithms still resemble the original description by
Metropolis~\cite{Metropolis53} in the fifties. For instance, Ising and other classical lattice models were for a very long time simulated using random spin flips. Such algorithms suffer from a critical slowing down in the neighborhood of the second order phase
transition. However, some twenty years ago Swendsen and Wang found a solution using cluster
updates~\cite{Swendsen87, Wolff89}, completely overcoming the critical slowing down for the classical Ising model. Monte Carlo methods have also been applied to
quantum many-body systems, where one tries to sample either the
wavefunction or the the partition function~\cite{Ceperley99}.
The quantum analog of cluster updates, namely the loop
algorithm~\cite{Evertz93}, triggered the development of the worm
algorithm~\cite{Prokofev98} and the operator loop algorithm~\cite{Sandvik99}
(and later the directed loop algorithm~\cite{Syljuasen02, Syljuasen03}) in the
stochastic series expansion representation. These algorithms share the
properties that they 
sample the one-body Green function, that they are formulated in continuous
time~\cite{Prokofev98} and are based on a world-line representation. Thus both algorithms are
similar~\cite{Troyer03}. These
algorithms have successfully been applied to spin systems and to the
Bose-Hubbard model~\cite{Kawashima03}. Recently, the worm algorithm has been formulated in the
canonical ensemble, allowing to study more systems including superconducting
grains and the nuclear pairing Hamiltonian~\cite{Rombouts05, Vanhoucke05}. 

The efficiency of a numerical simulation method is primordial: efficient
algorithms lead to more accurate results at the same computational cost and
allow for the study of larger systems. 
The algorithms mentioned above are based on a Markov process that results in a
random walk in a specific configuration space. Configurations are visited by
the random walker proportionally to their
respective weights. By the Markov process, subsequent measurements are trivially
correlated. The transition matrix specifies the probability of going from
one configuration to another, and has to be defined in advance. In practice, one
requires that transition matrices satisfy the principle of detailed
balance~\cite{Liu01}. This, however, still leaves some freedom in the choice of the
transition matrices, which can be used to optimize the efficiency of the
algorithm. A convenient updating scheme is the Metropolis-Hastings
algorithm~\cite{Metropolis53, Hastings70}: a limited number of
configurations can be reached from the current one by defining a proposal
distribution, and the transition to the new configuration is accepted or
rejected according to detailed balance~\cite{Liu01}. \\ 

In previous work, we introduced the notion of 'locally optimal
Monte Carlo'~\cite{Pollet04}, when the degrees of freedom of every transition
matrix are chosen in such a way as to mimic the globally optimal transition matrix,
which, at least in principle,  can be written down exactly. This approach is a
best guess for obtaining optimal efficiency, and was found successful in
practice for the directed loop algorithm in the stochastic series expansion
representation~\cite{Pollet04}. Here we investigate the consequences of the
locally optimal Monte Carlo idea for the worm algorithm, and try to combine
the advantages of the directed loop algorithm with the advantages of the worm
algorithm. This results in a new formulation of the worm algorithm, hereafter
called the locally optimal worm algorithm (LOWA)~\cite{Pollet05}. We show
results for various spin models and the Bose-Hubbard model, and compare the
efficiency of the LOWA with that of the directed loop algorithm in the
stochastic series expansion framework.     

\section{The algorithm}
We consider a two-body Hamiltonian $H$ defined on a discrete lattice of system size $L$.
The Hamiltonian can
be written as $H = H_0 - V$, where the terms $H_0$ and $V$ are to be specified
later on. We also assume a single particle basis $\vert i_\nu \rangle = \vert i_{\nu_1}, \ldots, i_{\nu_L} \rangle $ of $H_0$ such
that the action of any term in the Hamiltonian on a basis state yields a single
basis state. The models that we typically have in mind are the Bose-Hubbard
model, 
\begin{equation}
H = -t \sum_{\langle i,j \rangle} ^L{b}^{\dagger}_i{b}_j +
 \frac{U}{2}\sum_{i}^L{n}_i({n}_i-1) + V_{nn}\sum_{\langle i,j \rangle}^L n_i n_j 
 -\sum_{i}^L \mu {n}_i,
\end{equation}
and a general spin-$S$ model,
\begin{equation}
H = J\sum_{\langle i,j \rangle}^L \frac{1}{2}({S}^{+}_i{S}_j^{-} +
{S}^{-}_i{S}_j^{+}) + \Delta S_i^z S_j^z   -h \sum_i^L S_i^z.
\end{equation}
In both expressions the notation $\langle i,j \rangle$ refers to the sum over
nearest-neighbor sites only. In the Bose-Hubbard model, bosons are created on
site $j$ by the operator $b_j^{+}$ and the number of bosons on site $j$ is
counted by the number operator $n_j$. The kinetic term describes hopping of
the bosons with tunneling amplitude $t$, while we consider two kinds of
potential energy : on-site repulsion with 
strength $U$ and nearest-neighbor repulsion with strength $V_{nn}$. 
For the spin anti-ferromagnet (spin exchange amplitude $J>0$) , we require
that the lattice is bipartite. All matrix elements remain positive as long as
the model does not exhibit any frustration which can for instance be induced
by second nearest-neighbor hopping. As the calculations serve to demonstrate
the ideas related to efficiency, we restrict the discussion to one
dimension.\\   
 
A worm algorithm~\cite{Prokofev98} is a quantum Monte Carlo algorithm where
the decomposition of the partition function, 
\begin{eqnarray}\label{eq:zdecomp}
Z & = &  {\rm Tr} \sum_{n=0}^{\infty} \int_0^{\beta} dt_n \int_0^{t_n} dt_{n-1} \cdots
\int_0^{t_2} dt_1 \\ {} & {} & e^{-t_1 H_0}V
e^{-(t_2-t_1)H_0} 
\cdots e^{-(t_n - t_{n-1})H_0}V e^{-(\beta-t_n)H_0} \nonumber,
\end{eqnarray}
is sampled indirectly by performing local moves in the extended configuration space of open world-line configurations in the Green function sector,
\begin{equation}
Z_e = {\rm Tr} \left[ {\mathcal T} \left( \left(
  b_i(t_0)b_j^{\dagger}(\tau) + {\rm h.c.} \right) \exp(-\beta H) \right) \right].\label{eq:timeord}
\end{equation}
The symbol ${\mathcal T}$ denotes time ordering, and we have introduced the
Heisenberg operators $\left( b_i(t_0)b_j^{\dagger}(\tau) +
  b_i^{\dagger}(t_0)b_j(\tau) \right)$ which we call the 'worm' operators. Summing over all possible worms requires the extra sum evaluation of
 \begin{equation}
 \sum_{i,j,t_0, \tau} \ldots e^{-t_0H_0}b_i^{\dagger} e^{t_0H_0} \ldots e^{-t_{\tau}H_0}b_j e^{t_{\tau}H_0}\ldots,
 \end{equation}
where we have explicitly written out the Heisenberg worm operators, and where the '$\ldots$' mean that they should be inserted at the right time, as implied by the time ordering of eq.(\ref{eq:timeord}). 
Insertion of complete sets of basis states allows to replace the operators $H_0$ and $V$ by matrix elements. We will choose to work in the number occupation basis for the Bose-Hubbard model and in the spin $S_z$ basis for the spin models. A natural choice is to consider the one-body tunneling operators as perturbations $V$, and collect the diagonal one-body and two-body operators in $H_0$. This leads to the path-integral formulation. The stochastic series expansion representation is found back when all operators of the Hamiltonian are put into $V$ and $H_0=0$. The integration over time is then immediate. In the path-integral formulation, the extended partition function can be written as
\begin{equation}
Z_e = \sum_{n=0}^{\infty} \sum_{i_1, \ldots, i_n} \sum_{i,j} \int_0^{\beta} dt_n \int_0^{t_n} dt_{n-1} \cdots
\int_0^{t_2} dt_1 W,
\label{eq:ze}
\end{equation}
where the terms $W(n, t_1, \ldots, t_n, i,j, t_0,t_{\tau}, \vert i_{\nu} \rangle)$ can be interpreted as weights when positive. The weights depend on the order $n$ and the integration times $t_j, j = 1, \ldots n$ of the perturbative expansion of the partition function $Z$ in $V$ and also on the position and the time of the worm operators, and all possible inserted basis sets $\vert i_{\nu} \rangle $.
\begin{eqnarray}
W & = & \langle i_1 \vert V \vert i_2 \rangle e^{-(t_2-t_1)E_{i_2}} \langle i_2 \vert V \vert i_3 \rangle e^{-(t_3-t_2)E_{i_3}}  \ldots \nonumber \\
{} & {} & \ldots e^{-(t_k - t_{k-1}) E_{i_k}} \langle i_k \vert b_i^{\dagger} \vert i_ {k+1} \rangle \ldots  \nonumber \\
{} & {} & \ldots e^{-(t_l - t_{l-1}) E_{i_k}} \langle i_l \vert b_j \vert i_ {l+1} \rangle \ldots \nonumber \\
{} & {} & e^{-(t_n - t_{n-1}) E_{i_n}} \langle i_n \vert V \vert i_1 \rangle e^{-(\beta + t_1 - t_n)E_{i_1}}. \label{eq:weight} 
\end{eqnarray}
The Monte Carlo algorithm has to sample over all expansion orders $n$, all interaction times $t_i$, all possible worm times and positions, and all basis sets $\vert i_{\nu} \rangle$. In a worm algorithm this is achieved by moving one of the worm operators through the configuration space. Such updates shift, annihilate and create interactions and sample indirectly over all possible basis states.

\begin{figure}
\begin{center}
\includegraphics[width=8cm]{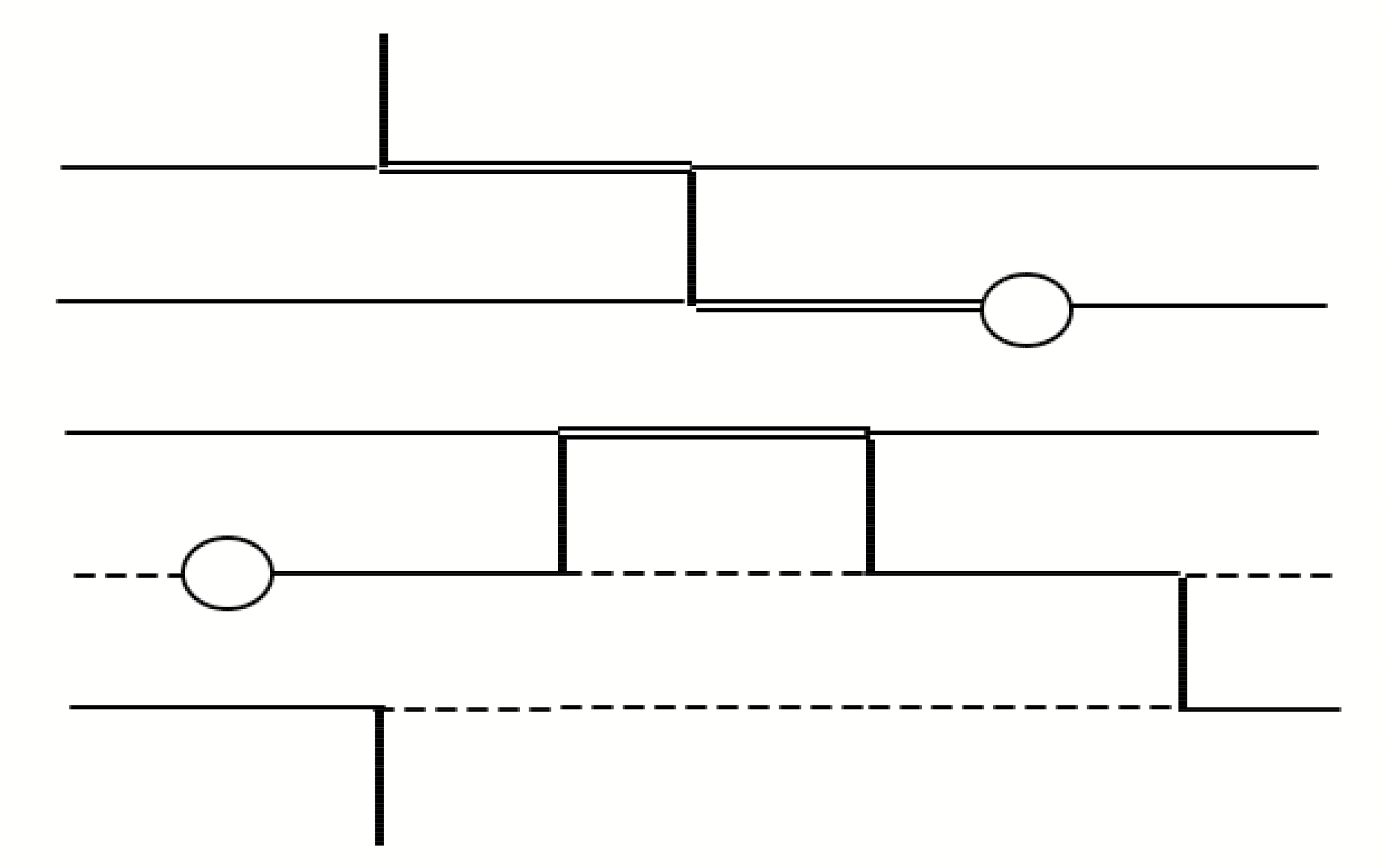}
\caption{\label{fig:conf_fig} Graphical representation of a typical configuration in the Green function sector. Time goes from left to right in the figure, there are five sites. World lines are denoted by single lines (site is once occupied), double lines (site has occupancy two) or dashed lines (site is not occupied). Interactions (hopping of a particle) are denoted by vertical lines. The two circles mark a discontinuity in the world lines and correspond to the worm operators. One of them creates an extra particle, the other one annihilates it. As a consequence of the $U(1)$ symmetry, total particle number is conserved at every interaction.}
\end{center}
\end{figure}

It is convenient to represent the weights in a graphical representation using world lines. An example is shown in Fig.~\ref{fig:conf_fig}.

Whenever the worm operators reach each other, the discontinuities in the graphical representation cancel and the resulting configuration belongs to the partition function $Z$, apart from the worm matrix element $\langle i_k \vert b_i(0)b_i^{\dagger}(0) \vert i_k \rangle$. At this point we are free to assign any value to this matrix element.  The standard choice, which we follow, is to take it constant for all states $\vert i_k \rangle$ (see eq.(\ref{eq:worm_weight_diag})). This is graphically represented by the absence of 'circles' (which denoted the discontinuities in Fig.~\ref{fig:conf_fig}) in a world-line picture.

In the literature there have been two different implementations of the worm idea. First, there was the 
worm algorithm by Prokof'ev {\it et al.}~\cite{Prokofev98}, which was formulated in the path-integral representation. Later the operator loop~\cite{Sandvik99} and directed loop algorithm~\cite{Syljuasen02, Syljuasen03} in the stochastic series expansion representation were formulated. 

Compared to the original formulation of the worm algorithm by Prokof'ev {\it et
  al.}~\cite{Prokofev98}, the efficiency of the directed loop algorithm in the
  stochastic series expansion representation was found superior for spin
  systems and even for a homogeneous Bose-Hubbard model in both phases (except
  for the extreme soft-core case)~\cite{Wessel04}. This
  is an expected result for spin systems where the diagonal energies are of
  the same order as the spin exchange amplitudes, but for the Bose-Hubbard
  model this result feels unsatisfactory. For a trapped Bose-Hubbard model
  however, the worm algorithm was found superior~\cite{Wessel04}. 

In the present paper we think of a new formulation of the worm idea, trying to combine to advantages of the directed loop algorithm with those of the worm algorithm. More specifically, we want an algorithm where the worm inserts and annihilates the interactions (as in the worm algorithm), but we also want to use the directed worms as in the directed loop algorithm. The modification of the diagonal factors and the hopping factors contributing to the weight, eq.(\ref{eq:weight}), should be done in a locally optimal way~\cite{Pollet04}.

\section{Updates and detailed balance}
The easiest way to ensure that the Markov chain converges 
to the correct invariant probability distribution is by assuring detailed balance\cite{Metropolis53},
\begin{equation}
W(X) T (X \to Y) q = W(Y) T(Y \to X).
\end{equation}
The current configuration is denoted by $X$, the new configuration by $Y$. 
$q$ is the acceptance factor to be used in the Metropolis algorithm,
where the transition rule for going from $X$ to $Y$ is given by
 $P (X \to Y)=T (X \to Y) \min(1,q) $, 
while the transition rule for the reverse update is given by
 $P (Y \to X)=T (Y \to X) \min(1,1/q)$, 
 such that $W(X) P(X \to Y) = W(Y) P(Y \to X)$.
We will now show that detailed balance is fulfilled for every possible update occurring in the algorithm.
 
\subsection{Move}
\begin{figure}[h]
\begin{center}
\includegraphics[width=8cm]{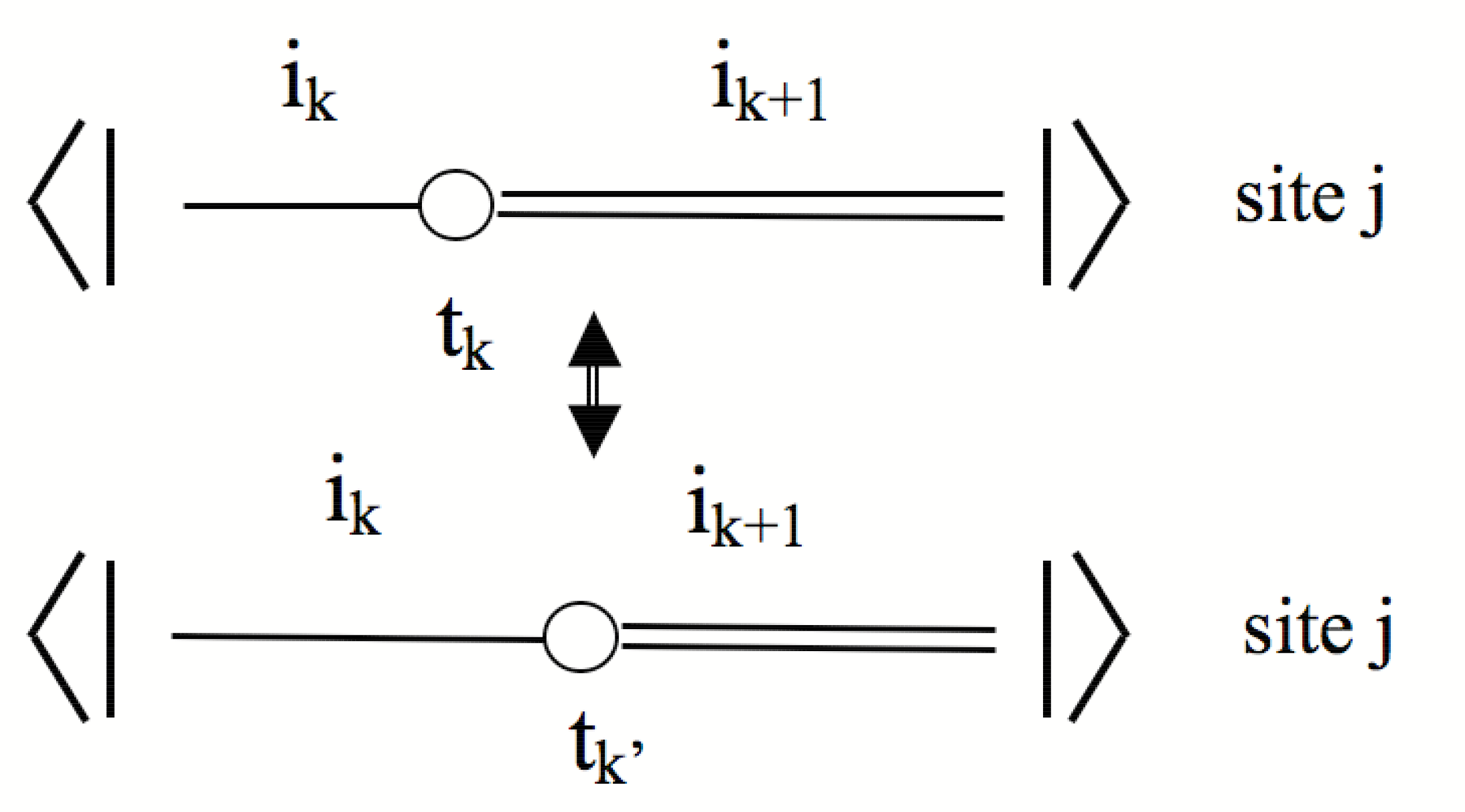}
\caption{\label{fig:move} Graphical illustration of the Move update. The worm, originally at time $t_k$ tries to jump to a later time $t_{k^\prime}$. The state to the left of the worm is $\vert i_k \rangle$, with energy $E_{i_k}$. The state to the right of the worm is $\vert i_{k+1} \rangle$, with energy $E_{i_{k+1}}$.}
\end{center}
\end{figure}

The simplest possible update is to move the mobile worm to another (later/earlier) time. We assume that the mobile worm is on site $j$, and corresponds to an operator $b_j$. The configurations before and after the update are shown in Fig.~\ref{fig:move}.

The relevant factors contributing to the weights before ($W(X)$) and after ($W(Y)$) the update are
\begin{eqnarray}
W(X) & = & e^{-t_k E_{i_k}} \langle i_k \vert b_j \vert i_{k+1} \rangle e^{t_k E_{i_{k+1}}} \\
W(Y) & = & e^{-t_{k^{\prime}} E_{i_k}} \langle i_k \vert b_j \vert i_{k+1} \rangle e^{t_{k^{\prime}} E_{i_{k+1}}}.
\end{eqnarray}
The notation $i_k$ means the complete state over all sites as in eq.(\ref{eq:weight}), but the subscript $j$ means the particular site $j$. The acceptance factor reads
\begin{equation}
q = \frac{e^{-\Delta t E_{i_k} }} {e^{-\Delta t E_{i_{k+1} }  }} \frac{T(Y \to X)}{T(X \to Y)},
\end{equation}
with $\Delta t = t_{k^{\prime}}  - t_k$. 
The exponentials can be canceled by choosing the transition probability densities as
\begin{eqnarray}
T(X \to Y) dt & = & E_{i_k} e^{-\Delta t E_{i_k}} dt \\
T(Y \to X) dt & = & E_{i_{k+1}} e^{-\Delta t E_{i_{k+1}} } dt. 
\end{eqnarray}
The normalization factors $E_{i_k}$ and $E_{i_{k+1}}$ enter into $q$. 
These factors will be taken into account explicitly 
together with the interaction matrix elements 
at the moment that interactions can be inserted/annihilated, see section~\ref{sec:insertandjump}.

Let $u$ be an uniform random deviate, $u \in [0,1[$. 
Then, $p = -\log u$ follows an exponential distribution 
and we can compute the time shift window $\Delta t$ 
as $\Delta t = p/E_{i_k} = - \log (u)/E_{i_k}$, with $E_{i_k} \ge 0$. 
We come back to this issue in Section~\ref{sec:refinements}. 
The recipe for the Move update is thus
\begin{enumerate}
\item Draw a random deviate $u \in [0,1[$.
\item When moving to the right (left), compute the time shift window $\Delta t = - \log (u) / E$, where $E$ is the diagonal energy to the left (right) of the worm.
\item If no interaction is encountered, update the current worm time from time $t$ to time $t + \Delta t$.
\end{enumerate}  
This amounts to a random walk based on Poisson steps.
The exponential factors contributing to the weight, eq.(\ref{eq:weight}), might fluctuate heavily. 
The strong point of the Poisson moves is that these exponential factors are canceled exactly.


\subsection{Inserting, removing and relinking an interaction}\label{sec:insertandjump}

So far we assumed that the worm did not encounter any interaction during its propagation. 
What happens if a worm does reach an interaction? 
At that point a decision must be taken:
can the worm pass the interaction or not? 
Or shall we delete the interaction? 
In that case we must also define updates to insert interactions 
that are in balance with the removal of interactions. 
In this section, we will first discuss two cases where the worm can pass the interaction, 
leaving it unchanged. 
Second, we discuss the insertion and removal of interactions which, as we will see, 
also incorporates the modification of interactions.


When the worm $b_i^{\dagger}$ encounters an interaction $b_k^{\dagger}b_l$ 
whose sites are different from the site on which the worm head resides ($ i \neq k,l $), 
the worm can pass the interaction with probability one. 
This is a consequence of the commutator of the worm operator 
and the interaction being zero, $[b_i^{\dagger}, b_k^{\dagger}b_l] = 0$~\cite{Pollet05}. 
Second, when a worm operator $b_i^{\dagger}$ encounters an interaction
$b_i^{\dagger}b_j$, it can pass the interaction with probability one. 
This is also the result of the commutator being zero, 
$[b_i^{\dagger}, b_i^{\dagger}b_j] = 0$~\cite{Pollet05}.

\begin{figure}[h]
\begin{center}
\includegraphics[width=8cm]{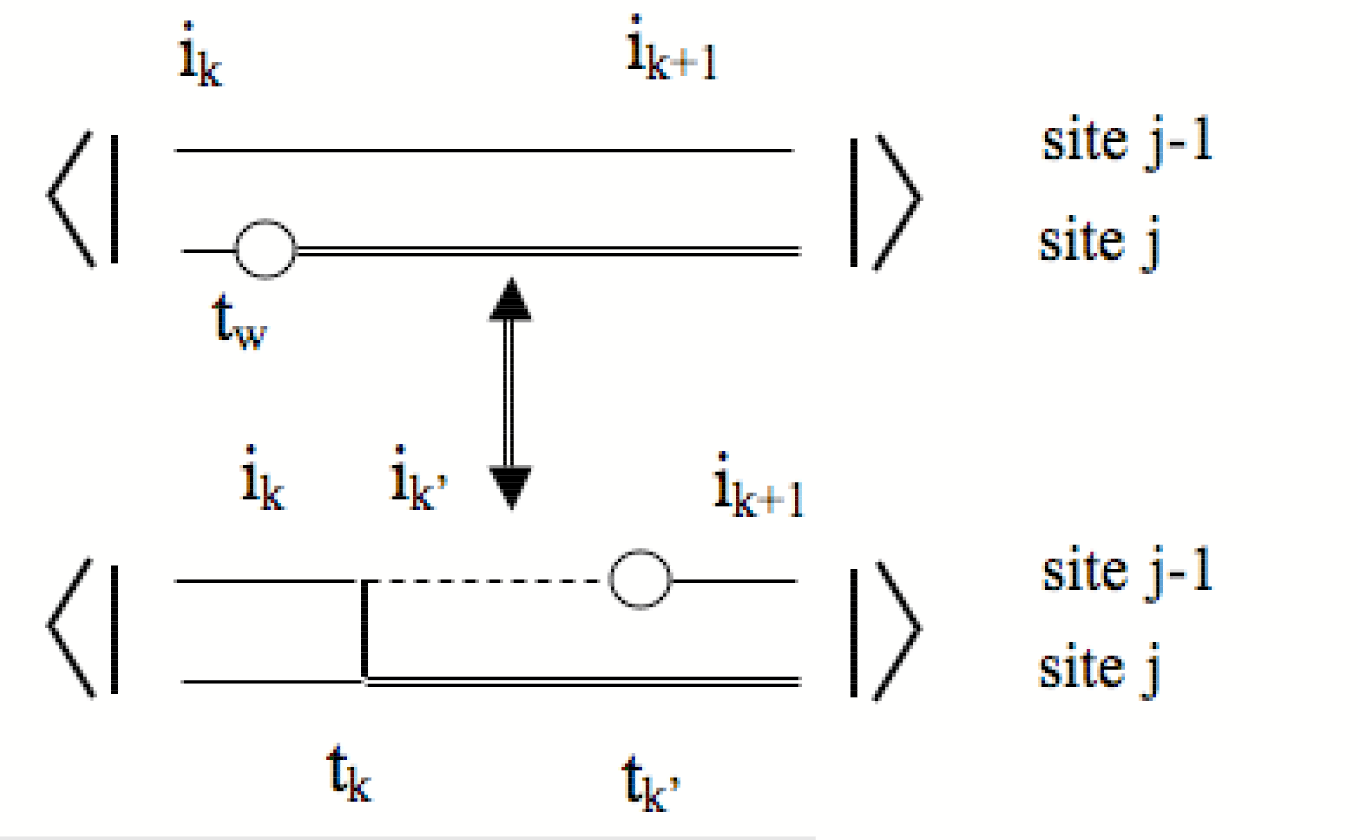}
\caption{\label{fig:interaction} Graphical illustration of the insertion of an interaction by the worm. The worm, originally at the time $t_w$ before the time $t_k$, jumps to the time $t_k$ and inserts an interaction there from site $j$ to site $j-1$. Afterwards the worm jumps to a later time  $t_{k^{\prime}}$, where we assume the worm always has to halt. Later we will relax this condition. Between times $t_k$ and $t_{k^{\prime}}$ a new state $i_{k^{\prime}}$ is created. Note that this update is only possible if the occupation on site $j-1$ is larger than zero.}
\end{center}
\end{figure}
Suppose we move to the right and want to insert an interaction. 
The update consists of a move from imaginary time $t_w$ to the time $t_k$ ,
inserting an interaction  
and then continue the move to the imaginary time $t_{k^{\prime}}$, as shown in Fig.~\ref{fig:interaction}.
Let's suppose that the worm pair was created at $t_w$, and that we want to evaluate estimators at time $t_{k^{\prime}}$, meaning that the worm has to halt at time $t_{k^{\prime}}$. We will later discuss generalizations.
The relevant contributions to the weights of the configurations before and after the update are
\begin{eqnarray}
W(X) & = & e^{-t_w E_{i_k}} \langle i_k \vert b_j \vert i_{k+1} \rangle e^{t_w E_{i_{k+1}}} \\
W(Y) & = & e^{-t_k E_{i_k}} \langle i_k \vert  b_j b_{j-1}^{\dagger} \vert i_{k^{\prime}} \rangle 
           e^{-(t_{k^{\prime}} - t_k)E_{i_{k^{\prime}}}} \nonumber \\
{} & {} &  \langle i_{k^{\prime}} \vert b_{j-1} \vert i_{k+1} \rangle e^{t_{k^{\prime}}E_{i_{k+1}}}.
\end{eqnarray}
The transition to move to the new configuration $Y$ is given by the probability density
\begin{equation}
T_M(X \to Y) dt = E_{i_{k}} e^{-\Delta t E_{i_{k}}}
                         e^{-\Delta t^{\prime} E_{i_{k^{\prime}}}} dt, \label{eq:tmxtoytransprob1}
\end{equation}
with $\Delta t = t_k - t_w$ and $\Delta t^{\prime} = t_{k^{\prime}} - t_k$.
The second exponential does not have an $E$ prefactor with it, since all time shifts larger than $\Delta t^{\prime}$ lead to the same final configuration $Y$. Similarly, for the reverse update
\begin{eqnarray}
T_M(Y \to X) & = & \int_{\Delta t}^{\infty} E_{i_{k+1}} e ^{-\tau E_{i_{k+1}}}d \tau  \times \nonumber \\
 {} & {} & \int_{\Delta t^{\prime}}^{\infty}  e^{-\tau E_{i_{k+1}}} E_{i_{k+1}} d\tau \nonumber \\
{} & = & e^{-\Delta t  E_{i_{k+1}}} e^{-\Delta t^{\prime} E_{i_{k+1}}}\nonumber \\
{} & = & e^{-(t_{k^{\prime}} - t_w) E_{i_{k+1}}}.
\end{eqnarray}
Once again, all the exponential factors cancel in the acceptance factor $q$. 
There remains a factor $\langle i_k \vert V \vert i_{k^{\prime}} \rangle / E_{i_k}$ (with $V$ the interaction),
which is taken into account in the equations of detailed balance 
for the actual insertion of a new interaction. If the worm was not forced to halt at the time $t_{k^{\prime}}$, the update would continue with the insertion of another interaction at $t_{k^{\prime}}$ (which would take the extra $E_{i_{k^{\prime}}}$ (i.e., the normalization factor of the second exponential in the right hand side of eq.(\ref{eq:tmxtoytransprob1})) into account).

\begin{figure}
\begin{center}
\includegraphics[width=8cm]{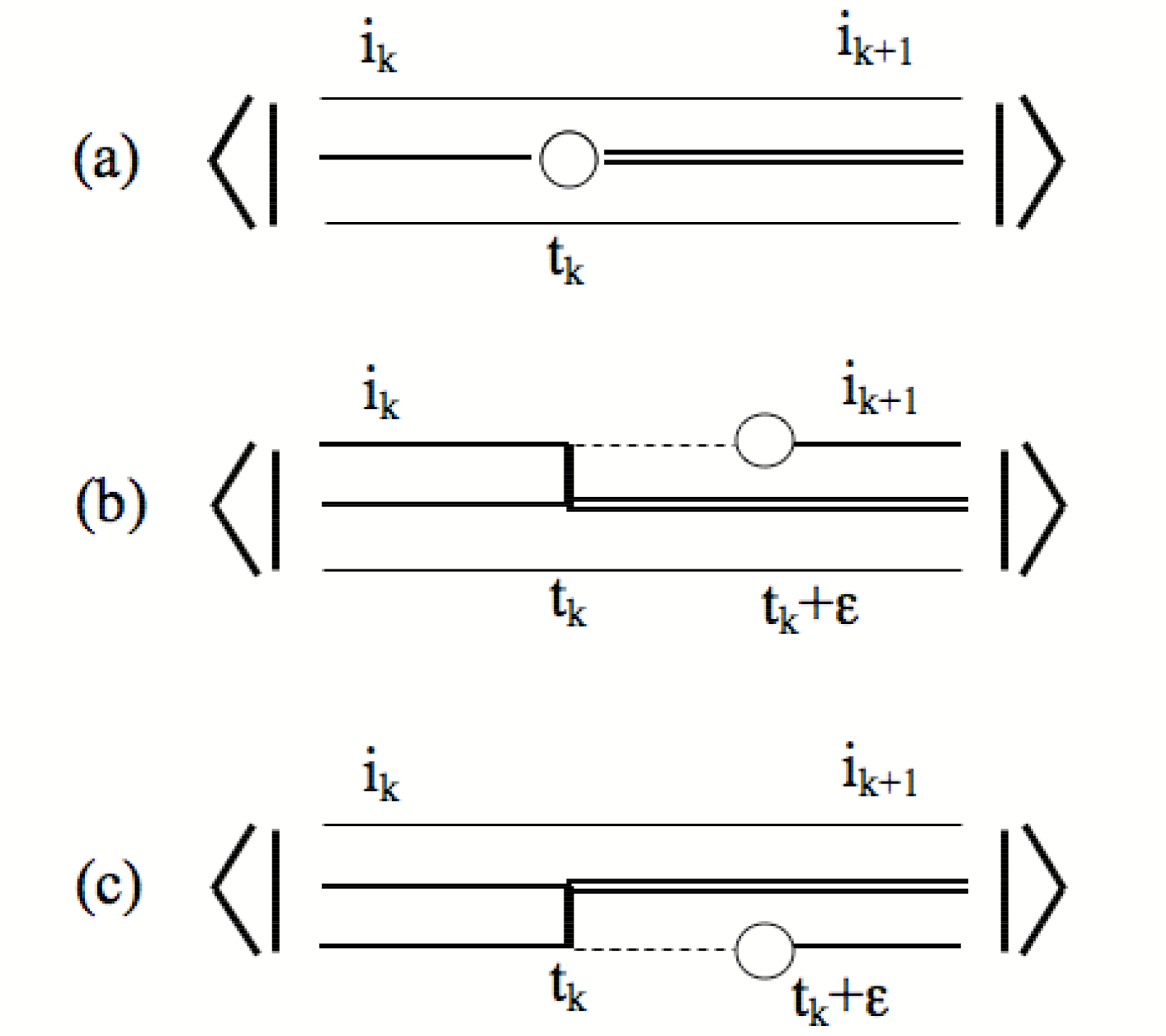}
\caption{\label{fig:diag_gworms} When the worm moves to the right and tries to
  insert an interaction in configuration (a), the two possible new
    configurations are configurations (b) and (c). The third possibility is
    bouncing back and changing direction in configuration (a). The transition
    matrix is thus a $3 \times 3$ matrix in one dimension, or a $(2d+1) \times (2d+1)$ matrix in $d$ dimensions.}
\end{center}
\end{figure}

When inserting a new interaction, we can either make the current site $j$ interact with site $j+1$ or site $j-1$ in one dimension. In higher dimensions $d$, interactions can be inserted to all $2d$ neighboring sites. 
We have to define the (conditional) probability distribution function that samples the three configurations of Fig.\ref{fig:diag_gworms}. When we are in configuration $a$, we have to define $P(a \to b)$ and $P(a \to c)$, both corresponding to the insertion of a new interaction. Similarly, when we are in configuration $b$ we have to define $P(b \to a)$ and $P(b \to c)$, corresponding to the removal and the modification (relinking) of an interaction, respectively. Updating between the configurations a, b and c should be done proportional to the following factors contributing to the weights,


\begin{eqnarray}
w(a) & = & \langle i_k \vert b_j \vert i_{k+1}  \rangle E_{i_k} \label{eq:weight_with_e} \\
w(b) & = & \langle i_k \vert b_j b_{j-1}^{\dagger} \vert i^{\prime} \rangle \langle i^{\prime} \vert b_{j-1} \vert i_{k+1} \rangle \\
w(c) & = & \langle i_k \vert b_j b_{j+1}^{\dagger} \vert i^{\prime \prime} \rangle \langle i^{\prime \prime} \vert b_{j+1} \vert i_{k+1} \rangle.
\end{eqnarray}
The energy term in eq.(\ref{eq:weight_with_e}) will be explained in section~\ref{sec:dirloopinvdistr}.
In Fig.~\ref{fig:diag_gworms}, the worm is on site $j$ in configuration (a), on site $j-1$ in configuration (b) and on site $j+1$ in configuration (c). We will discuss three possibilities to sample configurations (a), (b) and (c). 
\begin{itemize}
\item The 'Metropolis-like' way. Assume that we are in configuration (a). We choose with equal probability between (b) and (c). Say (b) was chosen. The transition to (b) is then accepted with probability $\min[1, w(b)/w(a)]$. If the update is not accepted, we stay in configuration (a). This approach has the advantage that only the matrix elements for configuration (b) and (a) have to be calculated or known, the ones for (c) are not needed. This approach is thus recommendable for high dimensions, and for long-range interactions. The (normalized) $3 \times 3$ transition probability matrix reads
\begin{equation}\label{eq:metropolis_def}
T^{\rm Met} = \left[
\begin{array}{ccc}
1-q_{ab}-q_{ac} & q_{ab} & q_{ac} \\
q_{ba}  & 1 - \ldots & q_{bc} \\
q_{ca} & q_{cb} & 1 - \ldots,
\end{array}
\right],
\end{equation}
with $q_{kl} = (1/2) \min[1, w(l)/w(k)]$. 
The diagonal element corresponding to the lowest weight is thus zero. 
\item The heat-bath way. We choose between (a), (b) and (c) according to their relative weights, $\pi(j) = w(j)/(\sum_k w(k))$, irrespective of the current configuration. The weights of all configurations are needed in order to evaluate their sum. The transition probability matrix is
\begin{equation}\label{eq:heatbath_def}
T^{\rm Hb} = \left[
\begin{array}{ccc}
\pi(a) & \pi(b) & \pi(c) \\
\pi(a) & \pi(b) & \pi(c) \\
\pi(a) & \pi(b) & \pi(c) \\
\end{array}
\right],
\end{equation}
\item The locally optimal way. Heat-bath updates can have relatively large diagonal elements, which are associated with large rejection ratios or large bounce probabilities. The principle of locally optimal Monte Carlo updating suggests an optimal transition probability matrix,
\begin{equation}\label{eq:loopt}
T^{\rm Lo} = \left[
\begin{array}{ccc}
0 & \frac{\pi_2}{1-\pi_1} & \frac{\pi_3}{1-\pi_1} \\
\frac{\pi_1}{1-\pi_1} & 0 & \frac{1-2\pi_1}{1-\pi_1} \\
\frac{\pi_1}{1-\pi_1} & \frac{\pi_2}{\pi_3}\frac{1-2\pi_1}{1-\pi_1} & 1-\ldots
\end{array}
\right],
\end{equation}
where the normalized weights $\pi(j)$ are now ordered in ascending order, $\pi_1 \le \pi_2 \le \pi_3$. The ordering of the weights makes sense only if they can be tabulated before the start of the actual simulation. The locally optimal matrix is the stochastic matrix with the lowest possible second largest eigenvalue, which is negative.
\end{itemize}
The non-zero diagonal elements correspond to bounces,
meaning that the worm head changes its direction and undoes
its changes until it reaches another point where an interaction
can be changed, removed or inserted.

\subsection{Insertion and Removal of a worm pair}

\begin{figure}
\begin{center}
\includegraphics[width=8cm]{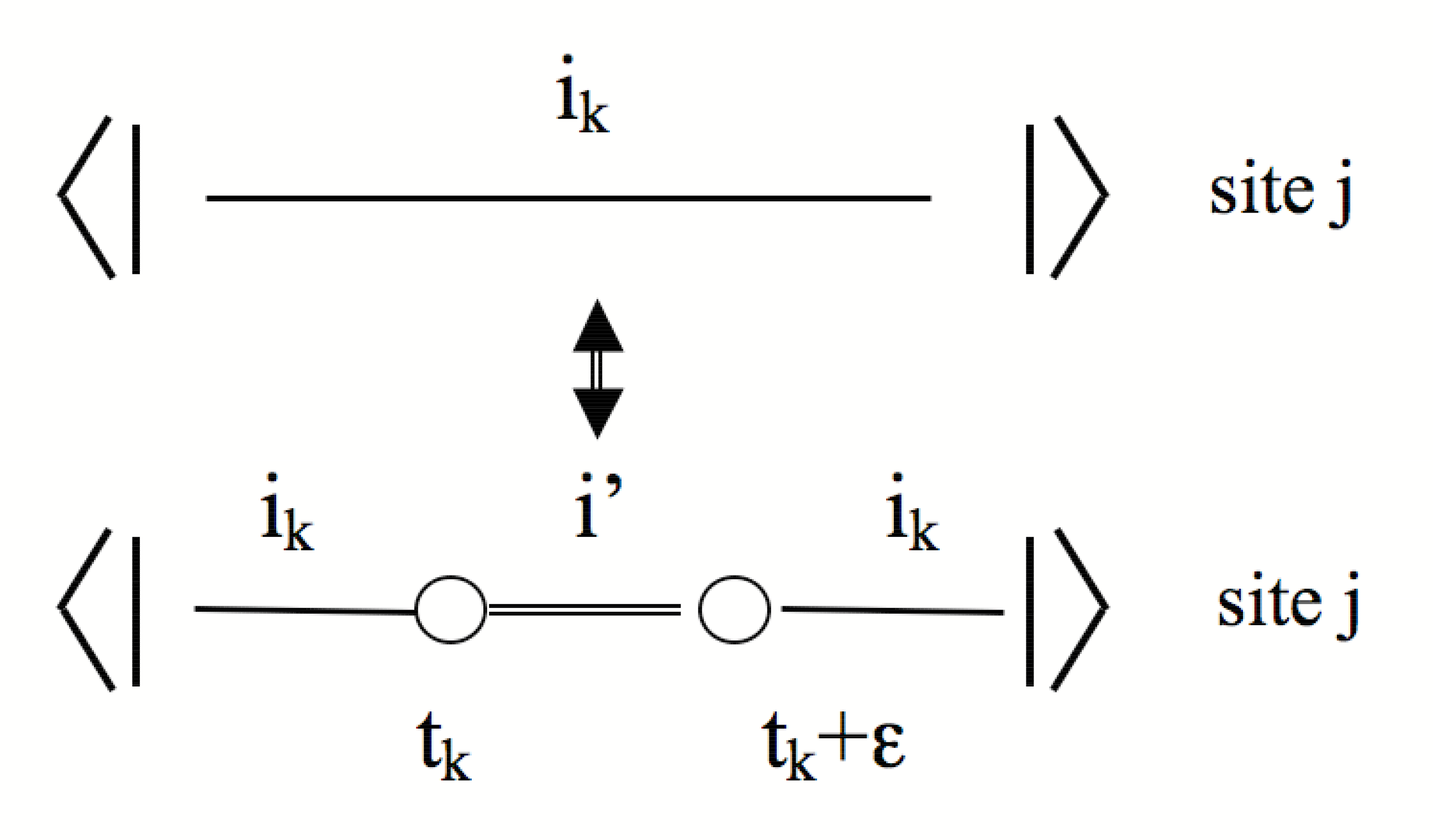}
\caption{\label{fig:insertion} Graphical illustration of the insertion of a worm pair. An arbitrary site $j$ and an an arbitrary time $t_k$ is chosen. We have shown here the case that the occupation between the worm ends is increased by one. }
\end{center}
\end{figure}

The only remaining update to discuss is the insertion/removal of a worm pair, which is the connection between the partition function sector $Z$ and the Green function sector, $Z_e$. The update is depicted in Fig.~\ref{fig:insertion}.

In case of Fig.~\ref{fig:insertion}, the weights before and after the update are
\begin{eqnarray}
W(X) & = & C_Z \label{eq:worm_weight_diag} \\
W(Y) & = & \langle i_k \vert b_j \vert i^{\prime} \rangle \langle i^{\prime} \vert b_j^{\dagger} \vert i_k \rangle  = (n_j+1),
\end{eqnarray}
where $C_Z$ is a constant specifying the relative weights of the partition function and the Green function sectors, and the number $n_j$ is the occupation on site $j$ before a worm pair is inserted. This is the usual convention in a worm algorithm. However, we are free to choose the weight of the worm matrix elements, and in SSE they are usually taken as unity. One of the worm ends will be moved through configuration space and is called the mobile worm, while the other end remains stationary.
We choose to always annihilate a worm pair when the mobile worm 'bites' into the stationary worm,
\begin{equation}
P(Y \to X) = 1.
\end{equation}
A worm pair is inserted by choosing a random time and a random site. Thus (in case of Fig.~\ref{fig:insertion}),
\begin{equation}
P(X \to Y) = \frac{1}{\beta}\frac{1}{L} P({\rm an}) P({\rm dir}).
\end{equation}
We have to define the probabilities $P({\rm cr})$ and $P({\rm an})$ that the occupation between the two worm operators is higher ('an') or lower ('cr') than outside this infinitesimal interval. We choose with equal probability among those, except when the occupation is zero or equal to the maximum occupation allowed. In case of zero occupancy, we choose with probability $50 \%$ not to insert a worm pair, and analogous to the case of maximum occupation number. Thus,
\begin{eqnarray}
P({\rm an}) & = & \frac{1-\delta_{n_j, N_{\rm Max}}}{2} \\
P({\rm cr}) & = & \frac{1-\delta_{n_j, 0                    }}{2}.
\end{eqnarray}
In case of hard-core bosons or spin$-1/2$ systems, one can modify this relation so as to always insert a worm pair. 

When a worm pair is inserted, there are two possibilities: of we move forward in time creating a particle, or we move backward in time creating a particle. Physically, moving forward in time and annihilating a particle is the same as the latter, while moving backward in time and annihilating a particle is the same as the former. Yet, the decorrelation benefits if both a direction (forward/backward) and an operation (creation/annihilation) are chosen. At this point, we explicitly choose a direction, and we take
\begin{equation}
P({\rm dir} = \rightarrow) = P({\rm dir} = \leftarrow) = 1/2. \label{eq:onehalfdir}
\end{equation}
This equation is correct in the directed loop algorithm since there are on average as many moves to the left as there are to the right.

Finally, we have to fix the constant $C_Z$. Apart from the cases $n_j=0$ and $n_j=N_{\rm Max}$, which we have already discussed, the update is always accepted when we choose (in case of Fig.~\ref{fig:insertion})
\begin{equation}
C_Z = 4 \beta L.
\end{equation}
With this choice, the Green function $G(x,t)$ is automatically correctly normalized. The procedure to measure will be discussed in Sec.~\ref{sec:dirloopinvdistr}, while the value of the measurements is the same as the ones discussed in the generalized directed loop algorithm~\cite{Alet03} and in the canonical worm algorithm~\cite{Rombouts05}.

\subsection{Refinements}\label{sec:refinements}
The perturbative expansion of the partition function is a Poisson process: the interactions (events) are distributed according to a Poisson distribution, with the intervals between them following an exponential distribution. We sampled these intervals exactly using exponential deviates divided by an energy. According to eq.(\ref{eq:weight}), these energies are the diagonal energies of the system. Taking into account that the direction of propagation is fixed in the directed loop algorithm, we only consider positive time shift windows $\Delta t$. Because the exponential distribution $\exp(-x)$ is only defined for positive $x$, the energies have to be positive, which is not a priori guaranteed. Let $E_{\rm L(R)} $ be the energy to the left (right) of the mobile worm. We can proceed in the following ways:
\begin{enumerate}
\item We shift all energies with a large, positive constant, $\epsilon_{\rm LR} = E_0 + E_{\rm LR}$. This will result in small time shift windows. Since these energies also enter in the equations of detailed balance for inserting/removing/relinking an interaction, they might lead to large bounce ratios.
\item We try to make use of the fact that only energy differences are physically relevant. We have the algorithmic freedom to choose any pair of $\epsilon_{\rm L}$ and $\epsilon_{\rm R}$ such that $\epsilon_{\rm L} - \epsilon_{\rm R} = E_{\rm L} - E_{\rm R}$ and $\epsilon_{\rm L}, \epsilon_{\rm R} \ge 0$. A good choice is  $\epsilon_{\rm L,R} = E_{\rm L,R} - \min \left[ E_{\rm L}, E_{\rm R} \right]$. This quantity can be zero,
  meaning that the jump in imaginary time is infinite, i.e., that the next
  interaction is always reached. Compared to the previous approach, we make thus larger jumps in imaginary time, and the energies that enter the detailed balance equations of inserting/removing/relinking an interaction are of the same order of magnitude as the hopping matrix elements. However, we found that this parameter choice resulted in some
  anomalously long (non-closing) loops and in problems with ergodicity when
  these long loops are discarded. 
\item We suggest to use the shifted energies 
\begin{equation}
\epsilon_{\rm L,R} = E_{\rm L,R} - \min \left[E_{\rm L}, E_{\rm R} \right] + E_{\rm off},\label{eq:shiften}
\end{equation}
 where the energy
  offset $E_{\rm off}$ overcomes the aforementioned problems with ergodicity. It
  makes sense to choose $E_{\rm off} = \langle V \rangle $ with $ \langle V
  \rangle$ a typical matrix element of the interaction.
\end{enumerate}
After the worm has passed an interaction at time $t_k$, we have to calculate the new
energy parameters $\epsilon_{\rm L,R}$ and consider a new time shift. The latter can be accomplished
by either drawing a new exponential deviate (as in step 3), or by adjusting the
time shift window $\Delta \tau \to \Delta \tau - (t_{k} - t_{k-1}) \epsilon_{\rm L}$
(when moving to the right), using the property that the exponential distribution has no 'memory'.

\subsection{Detailed balance and the directed loop algorithm}\label{sec:dirloopinvdistr}
To prove detailed balance one has to consider the global updates. 
Such a  global update starts and ends with a jump in imaginary time (thus only a forced halting at a chosen time can end a global update). In between, there can be any number  of insertions, annihilations or relinks. For example, the following sequence satisfies detailed balance : insertion of a worm pair - jump - insertion of an interaction - jump - insertion of an interaction - jump - removal of a worm pair.
We have already shown that the insertion and removal of a worm pair are balanced, so we just have to look at the sequence starting and ending with the jump.
For every such sequence there exists also the exact opposite sequence. 
Writing down the acceptance factor for the global move,
\begin{equation}
  q= \frac{W(Y) T(Y \to X)}{ W(X) T(X \to Y)},
\end{equation}
one finds that all exponential factors in the weights are cancelled out
by the probabilities for the Poisson jumps. 
For each insertion there enters an interaction matrix element
$\langle i_L \vert V \vert i_R \rangle$ in de numerator due to the ratio $W(Y)/W(X)$, 
and a factor $E$ ($E_L$ for moves to the right, $E_R$ for moves to the left)
in the denominator due to the normalization of the preceding Poisson jump. 
Both these factors are balanced through the locally optimal transition 
matrix of section~\ref{sec:insertandjump}. 
As a result one finds that $q$ is exactly equal to one,
hence all moves can be accepted, which greatly simplifies the computer code.
The resulting algorithm samples the configurations that appear in the decomposition
of eq.(\ref{eq:zdecomp}).
Note that the above reasoning still holds if we assume that the worm continues 
to move in the same direction after each Poisson move and only allow bounces
at the moment when one decides whether or not to insert or remove interactions.  
This leads to a version of the worm algorithm that is 
similar to the directed loop algorithm for SSE~\cite{Syljuasen02,Syljuasen03,Pollet05}.

The intermediate steps of the algorithm correspond to configurations
in the decomposition of the extended partition function $Z_e$ of eq.(\ref{eq:ze}).
To understand this, consider a point in imaginary time at a distance $\tau$
of the time where the worm was inserted.
Now suppose we {\it halt} (this is crucial for measuring the Green function) the worm head at the moment it passes the point $\tau$,
and suppose we choose with probability $1/2$ whether to continue the move in the
same direction or to move in the opposite direction 
(we will see furtheron that this latter condition is not necessary).
Following exactly the same reasoning as above, we see that this algorithm leads
to detailed balance between configurations in the decompositions of $Z$ and $Z_e(\tau)$.
Now from here we can derive that this is true even
if we continue the worm move in the same direction
at the moment of passing the point $\tau$.
To see this it is instructive to consider both directions
as two branches of a normalized transition kernel $\Psi$ that depends on two variables, 
namely the time $\Delta t$ and the direction $D$, and fulfills detailed balance:
\begin{equation}
  W(X) \Psi(\Delta t, \rightarrow) = W(Y) \Psi(\Delta t, \leftarrow),
\end{equation}
for $\Delta t = \vert t_Y - t_X \vert$ and $t_X < t_Y$.
We have that
\begin{eqnarray}
P(X \to Y) & = & \Psi(\Delta T, \rightarrow) \quad , X < Y \nonumber \\
{} & = & \Psi(\Delta T, \leftarrow) \quad , X > Y.
\end{eqnarray}
Thanks to time reversal symmetry, the statement that a worm creates a particle 
and propagates forward in time is completely equivalent to the the statement that 
the worm annihilates a particle and propagates backward in time. 
Therefore both directions will occur with equal probability.
Suppose that at a given moment a configuration $X$ is the actual one
with a probability proportional to its weight $W(X)$.
Then the probability for a configuration $Y$ to occur at the next step 
is proportional to
\begin{eqnarray}
\lefteqn{ \int W(X) P(X \to Y) dX} & &  \nonumber \\
{} & = & \sum_D \int  W(X) \Psi(\Delta t=\vert t_Y -t_X \vert, D) dX  \nonumber \\
{} & = &   \int_{t_Y>t_X} W(X) \Psi(\Delta t, \rightarrow) dX 
         + \int_{t_Y<t_X} W(X) \Psi(\Delta t, \leftarrow) dX \nonumber \\
{} & = &   \int_{t_Y > t_X} W(Y) \Psi(\Delta t, \leftarrow) dX 
         + \int_{t_Y<t_X}  W(Y) \Psi(\Delta t, \rightarrow) dX \nonumber \\
{} & = & \sum_D \int W(Y) \Psi(\Delta t,D) dX \nonumber \\
{} & = & W(Y) \int P(Y \to X) dX \nonumber \\
{} & = & W(Y).
\end{eqnarray}
This equation holds for any algorithm where the direction is fixed,
and also forms the basis of the {\em bounce algorithm} of ref.\cite{Pierleoni05}.

So, even when we do not change the direction of the worm at the moment 
that it passes the point $\tau$, 
the probability to pass a point $\tau$ is still proportional
to the weight of the corresponing configuration in the extended partition function $Z_e$.
This is quite subtle : Suppose we do a jump of the worm operator, without encountering an interaction. Then it is not possible to immediately come back to the original configuration in the next step since the direction is preserved. This observation lies at the heart of the proof of convergence of the algorithm given by the authors of Ref.~\cite{Syljuasen02, Syljuasen03, Alet03}. They prove that detailed balance is satisfied between any two diagonal configurations, from which it follows that every local step is balanced (in the SSE representation). Here, we see that detailed balance is fulfilled every time the worm is forced to halt at a chosen time, but we emphasize that precisely at the moment of inserting or annihilating or relinking an interaction (without further jump of the worm) detailed balance is not fulfilled.

Because the probability distribution for two consecutive Poisson steps in the same direction
is identical to the probability distribution for a single Poisson step,
one finds that in this case the dynamics of the algorithm is completely equivalent 
to the dynamics without considering a special point $\tau$.
Therefore one can state in general that the probability for the worm head to pass 
a point at a distance $\tau$ from the worm tail is given by the 
weight of the corresponing configuration in the extended partition function $Z_e$.

This observation allows for an efficient and unbiased evaluation of the equal-time
and unequal time Green function $G_{ij}(\tau)=Z_{e,ij}(\tau)/Z$:
each time the worm head at site $i$ passes the worm tail on a different site $j$,
one has a measurement for the equal time Green function $G_{ij}(0)$,
i.e., the one-body density matrix.
Counting the times that the worm passes at a distance $\tau$,
one obtains a measurement of the  unequal time Green function $G_{ij}(\tau)$.

\subsection{Stochastic Series Expansion representation}
As we have already mentioned, we end up in the Stochastic Series Expansion (SSE) representation if we treat all terms in the Hamiltonian as interactions. In the present formulation this means that the diagonal energies are always zero, leading to infinite time shift windows all the time. The mobile worm will jump from interaction to interaction, either deleting or relinking it or bouncing back. There is a serious problem with this algorithm, because it will never insert a new interaction. Therefore, in the SSE representation one needs two updates: one update is to scan over all (discrete) times in order to insert and remove interactions, the other one consists of modifying interactions with a fixed graph. The proof of convergence with directed loops in the second update proceeds in the same way as for the LOWA algorithm. 

\subsection{Summary of the LOWA algorithm}
We recapitulate and write down the full algorithm for the soft-core Bose-Hubbard model.
\begin{enumerate}
\item Pick an arbitrary site and an arbitrary time and call it $(i_0,  \tau_0)$. 
      Find the occupation on all sites at that particular time $\tau_0$. 
      Calculate the corresponding diagonal energy.  
      A direction (left or right) is chosen with equal probability. 
      Assume propagation to the right was chosen. 
\item At $(i_0, \tau_0)$ a worm-pair (tail-head) is inserted. 
      If the occupation is higher than zero, 
      the occupation between the worm ends can either be increased or decreased. 
      If the occupation at $(i_0, \tau_0)$ is zero, 
      then with probability $50 \%$ a  worm is inserted 
      with increased occupation between the worm ends. 
      With probability $50 \%$ no worm is inserted. 
\item When moving to the right (left), we denote by $\epsilon$ 
      the shifted energy to the left (right) of the worm head, eq.(\ref{eq:shiften}). 
      Draw an exponential deviate, $p = - \ln(u)$ with $u$ an uniform random number, $0<u \le 1$. 
      Evaluate the imaginary time shift window $\Delta t = p/\epsilon$ 
      and the new worm time $\tau^{\prime} = \tau + \Delta t$.
\item If the worm head encounters the worm tail (at the same site) during its propagation, 
      the update ends with probability one 
      and we arrive at a new diagonal configuration. 
\item If the new worm time is larger than the time to the next interaction,
      the new worm time only equals the time of the next interaction. 
      The worm can either bounce back, pass, annihilate or relink the interaction,
      according to the locally optimal transition matrix. 
\item If no interaction is encountered in the imaginary time shift window, 
      the worm shifts to its new time where an interaction is inserted or a bounce occurs,
      according to the locally optimal transition matrix. 
\item Go back to step 3.
\end{enumerate}

Every local single step respects the invariant distribution in the Green function sector.
When the mobile worm reaches the stationary worm, 
one can measure diagonal observables such as the energy, winding number, density, etc. 
They can be updated in the same way as in the directed loop~\cite{Syljuasen02} 
and worm~\cite{Prokofev98} algorithms.

The algorithm described above is valid when the diagonal energies involve
a single site, but also when the diagonal energies contain nearest-neighbor
repulsion repulsion terms, and even longer range interactions.

The worm algorithm in path-integral representation 
is in essence the same algorithm as the LOWA algorithm. 
The only difference is the way ergodicity and convergence 
to the correct invariant distribution are implemented 
with respect to the direction of worm propagation. 
In the worm algorithm, one chooses at every step 
between forward and backward propagation in time with equal probability, 
while in the LOWA the direction is maintained 
until an interaction forces the worm to alter its direction of propagation. 
In the context of the canonical worm algorithm, 
even other choices have been implemented~\cite{Rombouts05}. 
All these algorithms are the same in spirit, 
they are slightly different implementations of the same idea 
of performing local updates in the Green function sector. 
The directed loop algorithm has previously been
formulated in the path integral formulation~\cite{Syljuasen02}. 
Although there are some resemblances, the LOWA is different, more general, 
and the principles that lie at heart of the derivation of the algorithm are different.  

\section{Is the locally optimal worm algorithm efficient?}
Although we have argued why we believe the proposed LOWA is efficient,
we can only verify by doing numerics in order to get a definite answer on its
efficiency. The results are compared with data obtained by the directed loop
algorithm in the stochastic series expansion representation of
Refs.~\cite{Syljuasen02, Alet03, Pollet04}, which will be abbreviated as
DLSSE. We used the method of Ref.~\cite{Pollet04} for the actual computation of the DLSSE. 
The error and autocorrelation estimation was done using a binning analysis.

A direct comparison is complicated because present implementations use
different data structures for both algorithms. In the DLSSE,
a doubly linked list is constructed before the loop update. Since the graph is
fixed, the number of elements in this list cannot change. This allows to
allocate memory statically. In the worm algorithm on the other hand, the
number of interactions can change at any time. In our Fortran code this was
implemented using two arrays of a predetermined fixed length, corresponding to
the interactions before and after the current mobile worm time. 

\subsection{Bose-Hubbard model}
We have calculated the standard deviations on the kinetic energy and on the
squared density for a one-dimensional Bose-Hubbard model of size $L=32$ sites at an
inverse temperature of $\beta = 32$ and with a fixed chemical potential $\mu =
2$ in the absence of nearest-neighbor repulsion, $V_{nn}=0$. We work in units
$t=1$. Simulations consisted of 40 bins that each ran 300 seconds on a Pentium
III processor. We imposed a particle number cutoff of ten
particles per site for $U = 1, 2$ and $U=3$, while a cutoff of five particles
per site was taken for the other values of $U$, ranging from $U=4$ to
$U=10$. Imposing a cut-off is a necessity for the DLSSE, but
not for the LOWA. The number of loops per update
was optimized along the guidelines of Ref.~\cite{Alet03}. The Mott phase is
reached for $U=6$.
 
\begin{figure}[h]
\begin{center}
\includegraphics[width=9cm]{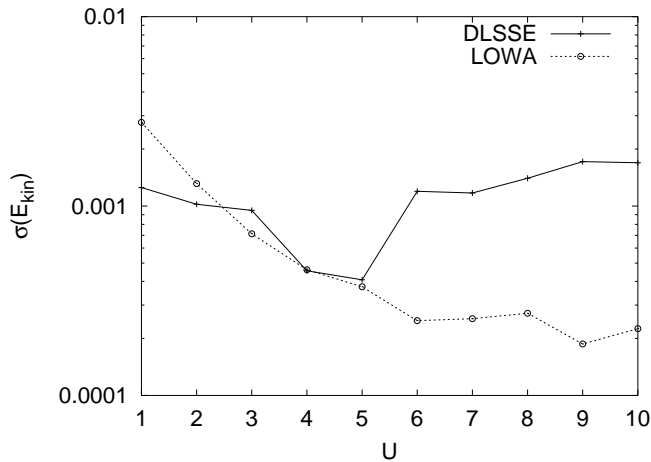}
\caption{\label{fig:gworm_ekin} Standard deviation (statistical error) on the
  kinetic energy for the directed loop algorithm in the stochastic series
  expansion framework (DLSSE), and the locally optimal worm (LOWA)  using an
  energy offset added to the energy difference for the diagonal energy
  parameters and 
  using locally optimal updates (eq.(~\ref{eq:loopt})). Simulations consisted
  of 40 bins that each ran 300 seconds on a Pentium III processor for a
  one-dimensional Bose-Hubbard model of $L = 32$ sites at an inverse
  temperature of $\beta = 32$ and with a fixed chemical potential $\mu =
  2$. The accuracy of the data points is about ten percent.}  
\end{center}
\end{figure}

\begin{figure}[h]
\begin{center}
\includegraphics[width=9cm]{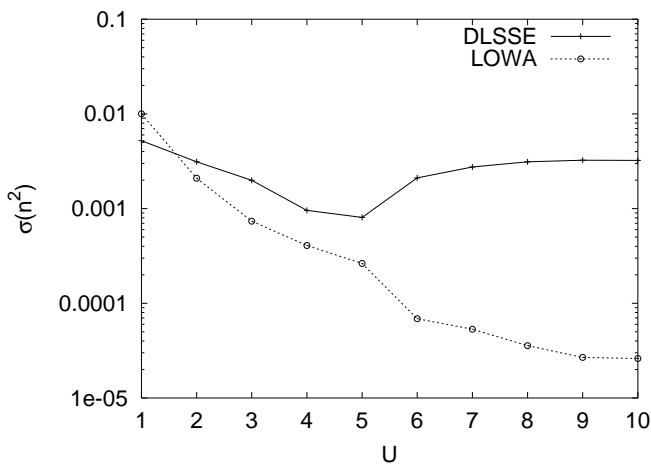}
\caption{\label{fig:gworm_dens}
  Idem as in Fig.~\ref{fig:gworm_ekin}, but now for the standard deviation on
  the average square density, $\langle n^2 \rangle = \langle (\sum_i n_i)^2 \rangle/L$.}
\end{center}
\end{figure}

We have calculated the standard deviation on the kinetic energy and on the
density squared (i) for the DLSSE. (ii) for the
LOWA where the diagonal energy parameters were chosen
according to approach (a) (iii) for the LOWA where the diagonal
energy parameters were chosen according to approach (c). In (ii) and (iii) the
$3 \times 3$ transition matrices were taken as the locally optimal ones as in
eq.(~\ref{eq:loopt}). We shall discuss this in section~\ref{sec:hb}. In
Figs.~\ref{fig:gworm_ekin} and~\ref{fig:gworm_dens} the results for algorithms
(i) and (iii) are shown.

Among the different LOWA optimization parameters, approach (iii) is
almost always the most efficient one. The DLSSE seems to be
the preferred model for very low values of $U$. However, approach (ii)
performs lots better than approach (iii) in this regime, and its efficiency is
comparable to that of the DLSSE. When the diagonal matrix
elements are much larger than the off-diagonal ones, as in the Mott phase
$U>6$, the present algorithm is superior. Admittedly, we recognize that it is
not unambiguous how the directed loop simulations should be performed in the
Mott phase because of the very short loop sizes.

\subsection{Spin systems} 
\begin{figure}[h]
\begin{center}
\includegraphics[width=9cm]{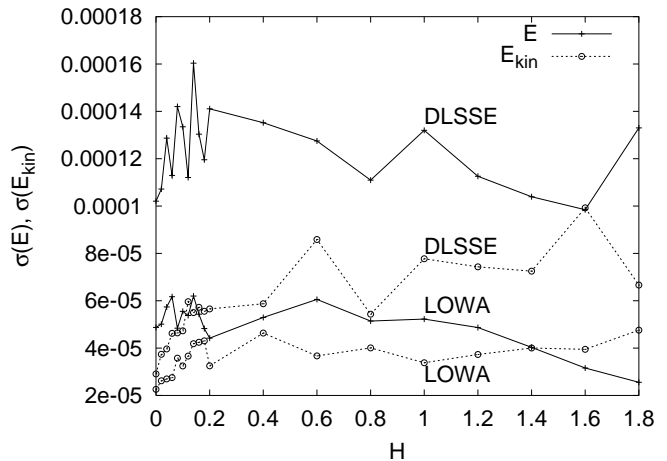}
\caption{\label{fig:fig_spin12_mag} Standard deviations on the total energy
  ($\sigma(E)$) and on the kinetic energy ($\sigma(E_{kin})$) obtained after a
  simulation of a spin-$1/2$ chain with $J = \Delta = 1$ consisting of $L =
  64$ sites and with inverse temperature $\beta = 16$. Simulations have been
  performed with the directed loop algorithm in the stochastic series
  expansion framework (DLSSE) and with the locally optimal worm algorithm
  (LOWA). Computations consisted of $40$ bins that each ran for $60$ seconds
  on a Pentium-III processor. }
\end{center}
\end{figure}

For spin systems, the magnitude of the diagonal and the off-diagonnal matrix
elements is of the same order. One can thus expect that the DLSSE is more
efficient for spin models than for soft-core bosonic models, 
since diagonal and off-diagonal operators are treated on equal footing in the
stochastic series expansion representation. It is thus interesting to compare the efficiency of the
locally optimal worm with the efficiency of the DLSSE for a
spin-$1/2$ chain. The LOWA was most efficient when the energy
offset parameter was set to $E_{\rm off} = 0.5$. In Fig.~\ref{fig:fig_spin12_mag}
the standard deviations of the total energy and of the kinetic
energy are shown, which are obtained by applying the LOWA and the
DLSSE to a spin-$1/2$ chain subject to a magnetic field
$H$. We find that the locally optimal worm is superior to the DLSSE in our
implementations, but it is more meaningful to say that both 
algorithms behave similarly as the magnetic field is increased, while the
algorithms are most  efficienct near $H=0$. Analogous conclusions were found
for a spin-$1$ chain. For a spin-$2$ chain however, the DLSSE was found to be
superior.

\subsection{Heat-bath updates and the locally optimal matrix}\label{sec:hb}

\begin{figure}[h]
\begin{center}
\includegraphics[width=9cm]{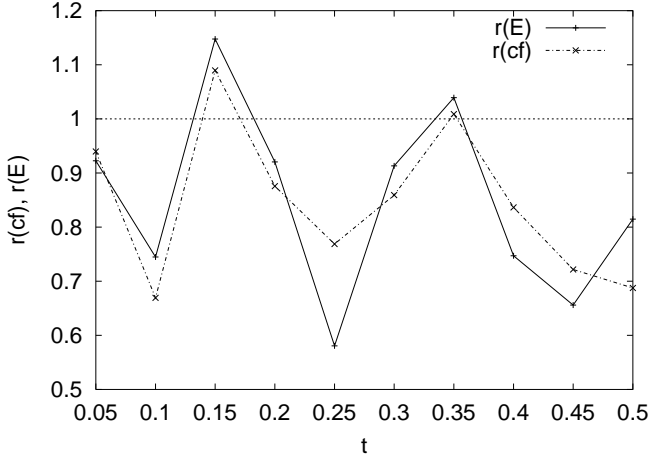}
\caption{\label{fig:hb_locopt} Simulation of a one-dimensional
Bose-Hubbard model with parameters $U=1, \mu = 0.5, \beta = 32, L = 32, V_{nn} =
0.4$ and varying tunneling amplitude $t$. Plotted is the ratio $r$ for the
condensate fraction $(r(cf))$ and for the total energy $(r(E))$, being the ratio
of the standard deviation $\sigma_{Lo}$ obtained by the locally optimal
approach to the standard deviation $\sigma_{Hb}$ obtained by the heat-bath
approach ($r = \frac{\sigma_{Hb}}{\sigma_{Lo}}$). Simulations consisted of 40
bins of 300 seconds on a Pentium III processor per data point.}
\end{center}
\end{figure}

A comparison between the heat-bath and
locally optimal approach is made in Fig.~\ref{fig:hb_locopt} for a
one-dimensional Bose-Hubbard model with parameters $U=1, \mu = 0.5, \beta =
32, L = 32, V_{nn} = 0.4$ and varying tunneling amplitude $t$. The ratio of
the standard deviation obtained by the locally optimal approach to the
standard deviation obtained by the heat-bath approach is shown for the
condensate fraction and the total energy. We see that the locally optimal
approach is on average ten to fifteen percent better, but the effect is less
pronounced than in the DLSSE~\cite{Syljuasen02, Syljuasen03, Alet03,
  Pollet04}. Even smaller differences were found for some other parameter
regimes.  

\subsection{Scaling of the worm with system size}
Both the worm and the DLSSE are ${\mathcal O}(N)$ methods. In the absence of correlations between subsequent measurements, the needed computation time for a desired accuracy scales linearly with system size and inverse temperature.
The scaling efficiency is further determined by the dynamical exponent $z$, which
describes how the integrated autocorrelation time scales with system size and
inverse temperature. The worm and directed loop algorithm have very low
dynamical exponents; $z$ is even zero in some high-dimensional cases. This
beneficient scaling is the cornerstone for the study of very large system
sizes at very low temperatures.
Since the present algorithm is based on the same principles as the directed
loop algorithm and the worm algorithm, one expects that the dynamical exponent
is similar (at least of the same order) but the prefactor of the scaling
behavior might be different.

We studied the scaling behavior for the critical system of an isotropic
spin-$1/2$ Heisenberg chain ($\Delta = J = 1$) in zero magnetic field ($H=0$), for
which the worm updates are fast. We
investigated the effects of increasing system size $L$ at fixed inverse
temperature $\beta$ on the one hand and of increasing the inverse temperature
$\beta$ at fixed system size $L$ on the other hand. This allows us to see
whether the algorithm scales symmetrically with system size and temperature or
not. All calculations ran for a fixed time of 40 bins of 300 seconds each on a
Pentium III processor. We used optimal parameters for the simulation: we use
the locally optimal transition matrix of eq.(\ref{eq:loopt}) and we set
$E_{\rm off} = 0.5$. We focused on the standard deviation and the integrated
autocorrelation time of the kinetic energy, since the modes of this observable
couple to the slowest mode of the simulation while the measurements can be
calculated at low computational cost.
\begin{figure}[h]
\begin{center}
\includegraphics[width=9cm]{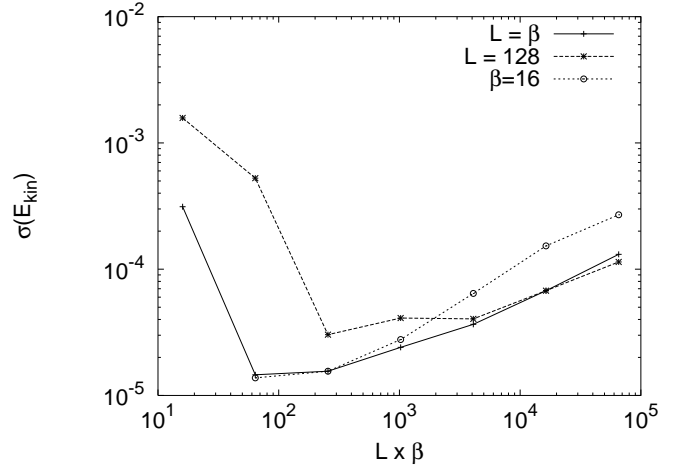}
\caption{\label{fig:scale_stdev} 
  Standard deviation on the average kinetic energy per site as a function of
  the system size $L$ multiplied by the inverse temperature $\beta$ for an
  isotropic spin-$1/2$ chain in zero magnetic field. Plots are shown when the
  system size and the inverse temperature are increased simultaneously ($L =
  \beta$), when the system size is held constant at $L = 128$ and only the
  temperature $\beta$ is varied ($L=129$), and when the temperature is held
  constant at $\beta = 16$ and the system size $L$ varies ($\beta = 16$).} 
\end{center}
\end{figure}
In Fig.~\ref{fig:scale_stdev} we study the standard deviation (statistical
error) on the average kinetic energy per site. When the system size and the
inverse temperature are sufficiently large, we see a gradual increase in the
statistical error with an exponent of $\sigma \sim (L\beta)^{0.5}$, irrespective
of whether $L$, $\beta$ or both (note that the quantum imaginary
time direction scales as one classical space direction) are increased.  Since, for larger lattices, the
worm will visit each site less often, this result is intuitively
understandable. From the data in Fig.~\ref{fig:scale_stdev} we can already see
that the dynamic exponent $z$ obeys $0 < z \le 1$. Strangely, when $\beta$ is
taken too low (see the data points at $L\beta = 16$ in
Fig.~\ref{fig:scale_stdev}), the present algorithm loses its efficiency. Below
we will relate this to higher integrated autocorrelation times.
\begin{figure}[h]
\begin{center}
\includegraphics[width=9cm]{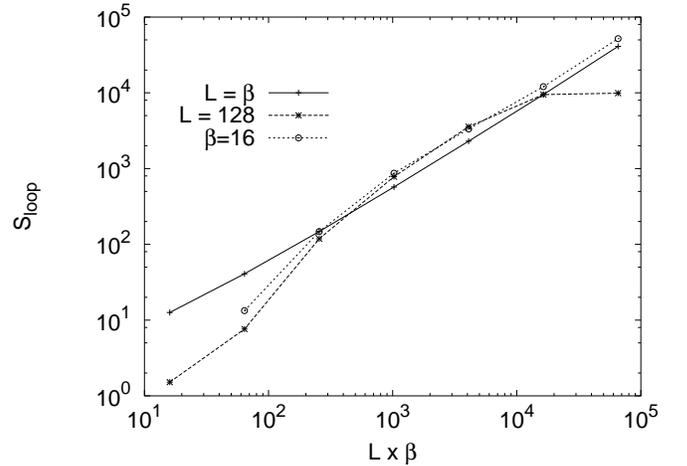}
\caption{\label{fig:scale_loopsize}
  Idem as in Fig.~\ref{fig:scale_stdev}, but now the loop size ($S_{\rm loop}$) is
  shown. The loop size $S_{\rm loop}$ is defined as the total number of
  interactions passed, inserted, annihilated and modified by the mobile worm
  in a single update.}  
\end{center}
\end{figure}
A similar picture results when we look at the loop size in
Fig.~\ref{fig:scale_loopsize}. We define the loop size $S_{\rm loop}$ as the total number of
interactions passed, inserted, annihilated and modified by the mobile worm in
a single update. We see that the loop size $S_{\rm loop}$ scales as $S_{\rm loop} \approx
(L\beta)^1$. The loop size increases thus linearly with the increase in area
in space-time. However, if we are very close to the ground state at a fixed
system size , then increasing $\beta$ does not result in longer loops and the
algorithm loses its scaling properties. 
\begin{figure}
\begin{center}
\includegraphics[width=9cm]{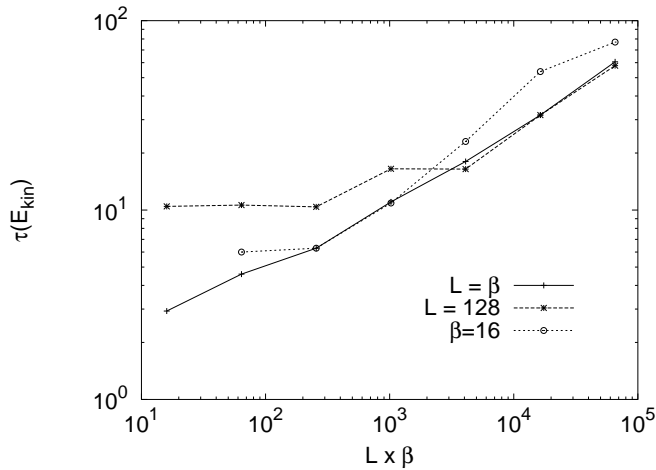}
\caption{\label{fig:scale_tau} 
  Idem as in Fig.~\ref{fig:scale_stdev}, but now the integrated
  autocorrelation time of the kinetic energy, $\tau(E_{kin})$ is shown.}
\end{center}
\end{figure}

The integrated autocorrelation time $\tau(E_{kin})$ of the kinetic energy in
Fig.~\ref{fig:scale_tau} shows a similar pattern: for large enough space-time
areas, the integrated autocorrelation time scales as $\tau(E_{kin}) \sim
(L\beta)^{0.42 \pm 2}$, which was fitted to the curve $L = \beta$ in Fig.~\ref{fig:scale_tau}.
When the inverse temperature is too low, we
unexpectedly find high autocorrelation times which explains the high standard
deviations for the same data points in Fig.~\ref{fig:scale_stdev}. This
behavior could be due to the fact that the mobile worm is always forced to
make relatively large jumps in the time direction.

With the DLSSE, the integrated autocorrelation time scales as $\tau(E_{kin})
\sim (L\beta)^{0.38 \pm 2}$. This result is in agreement with
Ref.~\cite{Syljuasen02}. The dynamic exponent of the DLSSE is
thus lower, yet in our calculations the standard deviations of the DLSSE increased more rapidly with increasing system size. This is due
to the increase in computational cost for a single update, which scales worse
for the DLSSE. The computational cost of a single update
depends strongly on the way of implementation, and the scaling of the standard
deviations with system size should be interpreted accordingly. In addition,
when looking at the magnetization on every site, the worm algorithm performs
much better than the DLSSE for all system sizes. We conclude
that efficiency largely depends on the implementation of the algorithm and on
the observables of interest~\cite{Alet03}.

\section{Conclusion}
In conclusion, we have presented a new formulation of the worm algorithm. 
The present algorithm has been derived using the concept of locally optimal \
Monte Carlo~\cite{Pollet04} and incorporates ideas both from the worm
algorithm~\cite{Prokofev98} and the directed loop algorithm~\cite{Syljuasen02}
in the stochastic series expansion representation~\cite{Sandvik99}. 
We have compared the efficiency of the present algorithm with that 
of the directed loop algorithm for spin chains and for the Bose-Hubbard chain. 
Especially when there are large diagonal matrix elements, 
the present worm algorithm is very successful. 
We have shown that choosing the locally optimal matrix for the transition matrices 
occurring in the stochastic subprocesses yields an efficient algorithm. 
We found that the loop size increases linearly with the increase in area in space-time, 
and that the dynamic exponent equals $z=0.84 \pm 4$ for an isotropic Heisenberg chain 
without magnetic field. 
Seen the efficiency of the method and its advantageous scaling properties, 
the algorithm is suitable for large scale calculations of spin systems 
and soft-core bosonic models.


The authors wish to thank the Research Board of the Universiteit Gent and the
Fund for Scientific Research, Flanders for financial support. 
The authors acknowledge D. Ceperley, E. Gull, N. V. Prokof'ev, B. V. Svistunov, M. Troyer, S. Wessel 
and F. F. Assaad for valuable discussions.

\end{document}